\documentclass[twocolumn,appendixfloats]{aastex6}
\slugcomment{{\sc Accepted to AJ:} 03/07/2017}
\usepackage{graphicx}
\usepackage{natbib}
\usepackage{latexsym}
\usepackage{amssymb}
\usepackage{longtable}
\usepackage{amsmath}
\usepackage{url}
\def\msun{\ifmmode {\rm\,M_\odot}\else ${\rm\,M_\odot}$\fi}
\def\Msun{\ifmmode {\rm\,\it{M_\odot}}\else ${\rm\,M_\odot}$\fi}
\def\lsun{\ifmmode {\rm\,L_\odot}\else ${\rm\,L_\odot}$\fi}
\def\Lsun{\ifmmode {\rm\,\it{L_\odot}}\else ${\rm\,L_\odot}$\fi}
\def\rsun{\ifmmode {\rm\,R_\odot}\else ${\rm\,R_\odot}$\fi}
\def\Rsun{\ifmmode {\rm\,\it{R_\odot}}\else ${\rm\,R_\odot}$\fi}
\def\Tsun{\ifmmode {\rm\,T_\odot}\else ${\rm\,T_\odot}$\fi}
\def\arcsec{\ifmmode {^{\prime\prime}}\else $^{\prime\prime}$\fi}
\def\asec{\ifmmode {^{\prime\prime}}\else $^{\prime\prime}$\fi}
\def\arcmin{\ifmmode {^{\prime}}\else $^{\prime}$\fi}
\def\amin{\ifmmode {^{\prime}}\else $^{\prime}$\fi}
\def\simlt{\mathrel{\spose{\lower 3pt\hbox{$\mathchar"218$}}\raise 2.0pt\hbox{$\mathchar"13C$}}}
\def\simgt{\mathrel{\spose{\lower 3pt\hbox{$\mathchar"218$}}\raise 2.0pt\hbox{$\mathchar"13E$}}}

\begin{document}

\title{Evidence for abnormal H$\alpha$ variability during near-transit observations of HD 189733 b}

\author{P. Wilson Cauley and Seth Redfield}
\email{pcauley@wesleyan.edu}
\affil{Wesleyan University\\
Astronomy Department, Van Vleck Observatory, 96 Foss Hill Dr., Middletown, CT 06459}

\author{Adam G. Jensen}
\affil{University of Nebraska-Kearney\\
Department of Physics \& Astronomy, 24011 11th Avenue, Kearney, NE 68849}

\begin{abstract} 

Changes in levels of stellar activity can mimic absorption signatures in transmission spectra from circumplanetary material. The frequency
and magnitude of these changes is thus important to understand in order to attribute any particular signal
to the circumplanetary environment. We present short-cadence, high-resolution out-of-transit H$\alpha$ spectra
for the hot Jupiter host HD 189733 in order to establish the frequency and magnitude of intrinsic stellar variations in the H$\alpha$ line core. We find that
changes in the line core strength similar to those observed immediately pre- and post-transit in two independent data sets
are uncommon. This suggests that the observed near-transit signatures are either due to absorbing circumplanetary
material or occur preferentially in time very near planetary transits. In either case, the evidence for abnormal H$\alpha$
variability is strengthened, although the short-cadence out-of-transit data do not argue for circumplanetary
absorption versus stellar activity caused by a star-planet interaction. Further out-of-transit monitoring at higher signal-to-noise would be
useful to more strictly constrain the frequency of the near-transit changes in the H$\alpha$ line core. 

\end{abstract}


\section{INTRODUCTION}
\label{sec:intro}

Much effort has been invested into understanding how stellar activity affects radial velocity
measurements and broadband transit observations \citep[e.g.,
][]{saar97,pont11,aigrain12,dumusque14,oshagh14,andersen15,llama15,herrero16,hebrard16,chiavassa17} and many
studies exist of long-term activity variations, i.e. day to year timescales, for known exoplanet
hosts \citep{boisse09,fares10,gomes11,gomes14,figueira16,giguere16}. On the other hand, little
investigation has been aimed at understanding stellar activity variations on very short timescales
for known exoplanet host stars (i.e., minutes to hours). This is unsurprising as long-term variations play
a more significant role in RV data sets collected sporadically across many nights.
Similarly, even large variations in spectroscopic activity indicators (e.g., H$\alpha$, \ion{Ca}{2},
or \ion{Na}{1}) are completely washed out in broadband photometric observations.  With the
increasing popularity of short-cadence, high-spectral resolution transit observations
\citep[e.g.,][]{redfield,jensen12,wyttenbach,cauley15,cauley16,barnes16} it is important to
understand the short-term behavior of host stars in order to differentiate between true absorption
by planetary material and stochastic changes in stellar activity indicators potentially caused by star-planet
interactions (SPIs).

In \citet{cauley15,cauley16} we reported on pre- and post-transit H$\alpha$ absorption
signatures in high-resolution data taken across two different transits of HD 189733 b. Similar
signatures have been reported in atomic UV transitions for HD 189733 b \citep{benjaffel,bourrier13}
and WASP-12 b \citep{fossati,haswell12}.  Each of the H$\alpha$ signatures showed a different depth and
time-series shape, suggesting that the physical mechanism is highly variable. Since absorption
measurements from high-resolution transmission spectra are necessarily normalized relative to a
reference spectrum, or a group of spectra, the absorption is by definition relative to another point
in time. Thus if the reference spectrum is chosen during a time of higher stellar activity, other
spectra will show absorption, mimicking the loss of photons by line-of-sight material. The
short baselines, i.e., a single night, for the high-cadence transit observations make it difficult
to distinguish between these changes in the stellar activity level and true absorption by
circumplanetary material.

In this paper we present out-of-transit observations of HD 189733 in order to further probe the frequency of
the observed pre- and post-transit changes in the H$\alpha$ line core. \autoref{sec:observations} describes the observations and data
reduction procedures. In \autoref{sec:timeseries} we present the H$\alpha$ time series data and
discuss the statistical analysis of the changes in the H$\alpha$ signal. A
brief summary is given in \autoref{sec:summary}.

\section{OBSERVATIONS AND DATA REDUCTION}
\label{sec:observations}

\begin{figure}[t]
   \includegraphics[scale=.57,clip,trim=57mm 30mm 30mm 50mm,angle=0]{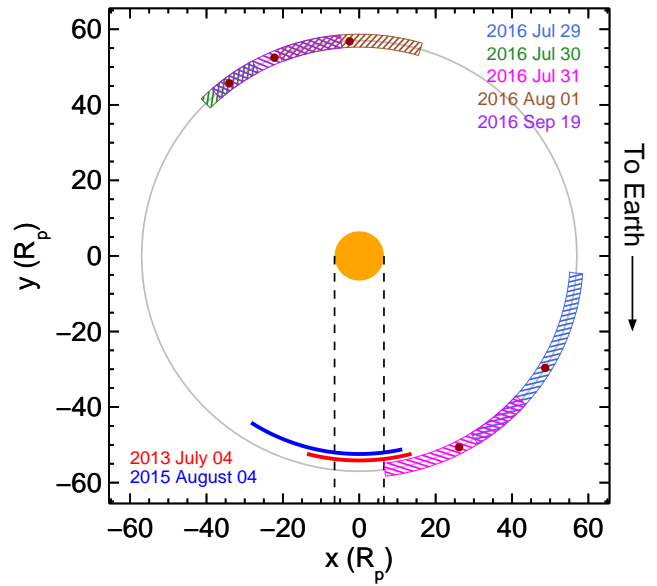} 
   \figcaption{To-scale diagram showing the orbital phases during which the 2016 McDonald observations were performed
   (hatched regions) and the transit observations from \citet{cauley15,cauley16} (solid red and blue lines). The in-transit portion
   of the orbit is marked with vertical dashed lines and the mean planet position during each set of 2016 observations is marked
   with a brown circle. \label{fig:phasevis}}
\end{figure}

Observations were performed using the Tull Coud{\'e} Spectrograph \citep{tull95} on the Harlan J. Smith
2.7 meter telescope at McDonald Observatory. We collected data on five separate nights. A to-scale
diagram of the phases during which the observations took place is show in \autoref{fig:phasevis}. The night
of 2016 July 29 was shortened due to persistent high humidity and 2016 September 19 was shortened because
of heavy clouds during the first two hours of the night. The nights of 2016 July 31 and 2016 August 01
were shortened due to observatory functions requiring use of the telescope during the first third of each night.
Details concerning the 2013 and 2015 Keck observations are presented in \citet{cauley15,cauley16}.

Spectra were taken with two different slits. We utilized the \#5 slit for the first night of 2016 July 29 which has 
a width of 1.79$^{\prime\prime}$ and resolving power of $R \sim 40,000$, or $\sim$7.5 km s$^{-1}$. These observations were
typically 900 seconds in length. For the nights of 2016 July 30 and July 31 we used the \#6 slit with $R \sim 30,000$
(or $\sim$10 km s$^{-1}$) and 600 second integrations in order to use a more similar
cadence to the past Keck observations and to boost the signal to noise. Exposure times were increased to 900 s on 2016 August 01 due
to moderate cirrus cloud coverage. The telluric standard HR 8634 was observed for approximately ten minutes each night.
The spectra from 2016 July 29 were broadened to match the instrumental resolution of the observations
taken with the \#6 slit in order to standardize the data collected with different spectrograph settings.

Standard data reduction steps were taken, including bias subtraction, flat fielding with a median
flat, and optimal extraction using custom IDL routines. Spectra were extracted for two separate orders,
one containing the H$\alpha$ line and another containing a comparison \ion{Fe}{1} line at 6677.98 \AA.
The \ion{Fe}{1} line is used as a control to ensure that the extraction and spectrum comparison
procedures are not artificially producing signals in the line cores.
Wavelength solutions for the individual orders
were found using Th-Ar lamp exposures and 3rd or 4th degree polynomial fits. Typically $\sim$20-30 Th-Ar
lines were used in the solution. Small wavelength shifts between individual spectra are corrected using 
strong stellar lines in the same order besides the line of interest. Each wavelength solution is then corrected
for the barycentric velocity and HD 189733 system radial velocity, which we take to be
$-2.23$ km s$^{-1}$ \citep{digloria15}.

The telluric spectrum in the H$\alpha$ order was modeled using the program Molecfit \citep{kausch}.
We modeled the telluric spectrum in the \ion{Fe}{1} order but it is very weak and results
in negligible changes to the transmission spectrum.
We first remove the blaze function and the broad stellar H$\alpha$ line from the telluric standard
spectrum using a high order spline. Molecfit then fits the normalized telluric spectrum for the
H$_2$O and O$_2$ column densities, as well as the instrumental resolution. This master telluric
model is then scaled, shifted, and divided out of the individual observations.  

\begin{figure*}[t]
   \centering
   \includegraphics[scale=.73,clip,trim=8mm 20mm 5mm 18mm,angle=0]{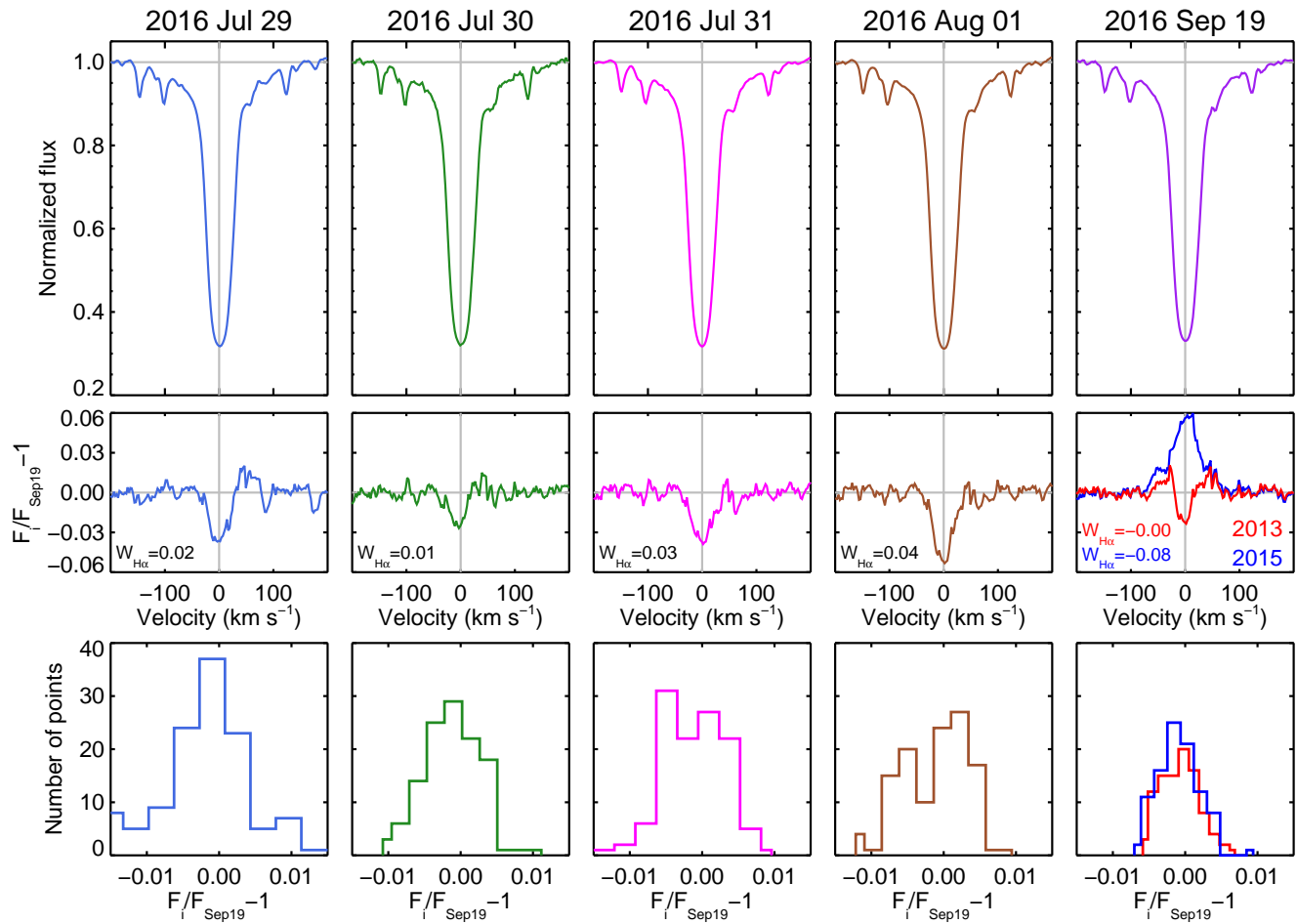} 
   \figcaption{Average H$\alpha$ spectra for each night (top panel) and the ``transmission'' spectrum (see \autoref{eq:strans})
   relative to 2016 September 19 (middle panels). The spikes near $\sim$90 and 190 km s$^{-1}$ in the
   2016 July 29 $S_T$ spectrum are telluric residuals. The equivalent width of the ratio spectrum, $W_{H\alpha}$ 
   (also see \autoref{eq:wlambda}), is given in the middle panels. All of the July nights show less core emission
   than 2016 September 19, suggesting that HD 189733 was in a more active state during 
   the 2016 September 19 observations. The 2015 H$\alpha$ core (bottom right panel) was strongly
   filled in relative to the 2016 observations, confirming the high activity state noted in \citet{cauley16}. The bottom
   panels show histograms of the transmission spectrum for $50$ km s$^{-1}$ $< |v| < 200$ km s$^{-1}$. The
   mild departure from purely Gaussian noise in the transmission spectra can be attributed to imperfect
   telluric subtraction. \label{fig:haspecs}}
\end{figure*}

\begin{figure*}[t]
   \centering
   \includegraphics[scale=.73,clip,trim=8mm 20mm 5mm 18mm,angle=0]{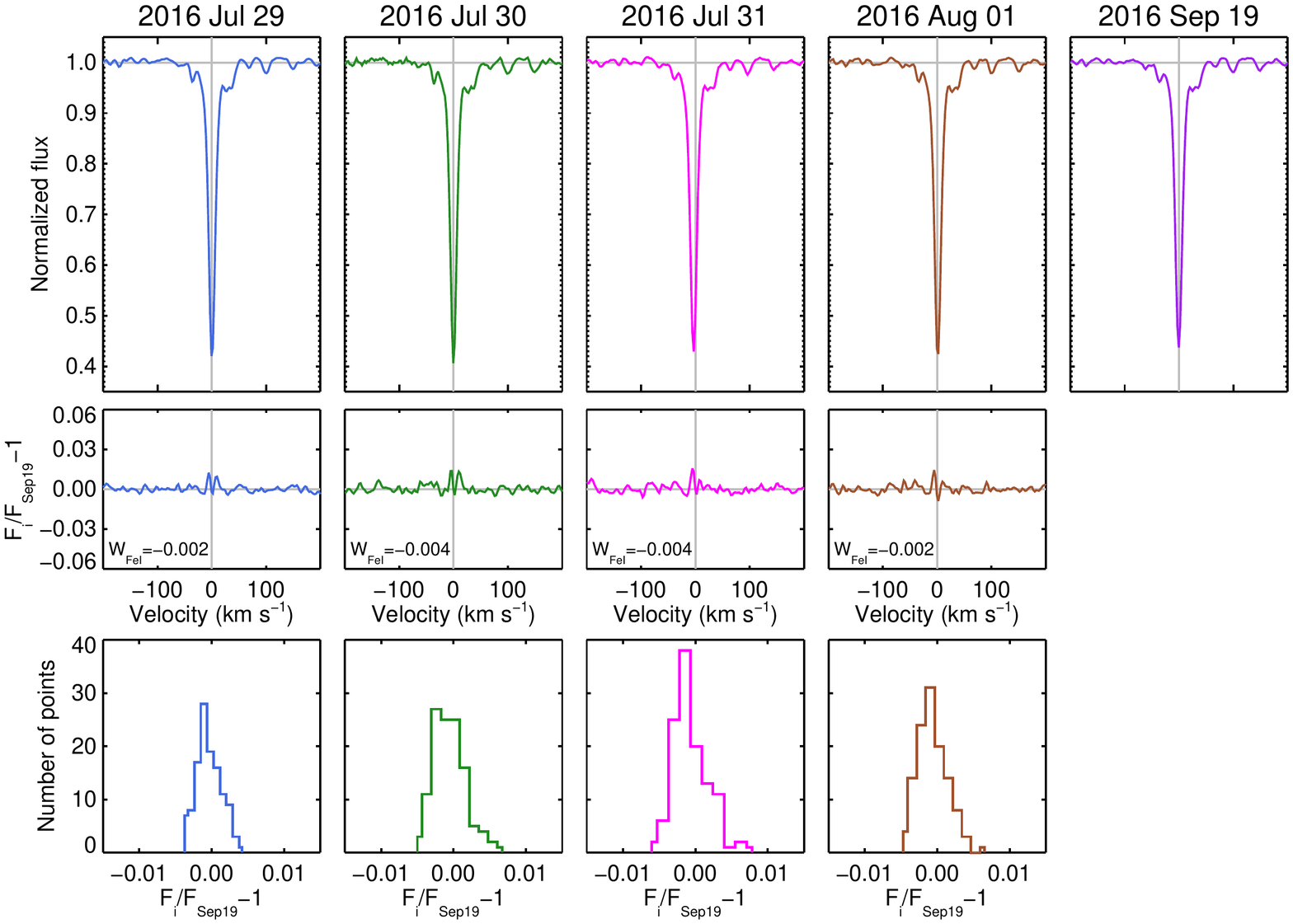} 
   \figcaption{Same as \autoref{fig:haspecs} but for the \ion{Fe}{1} 6677.98 \AA\ control line. The \ion{Fe}{1} line
   is less sensitive to changes in stellar activity levels and the observed differences compared to the
   2016 September 19 spectrum are negligible. \label{fig:feispecs}}
\end{figure*}

The mean normalized H$\alpha$ spectrum from each night is shown in the top row of
\autoref{fig:haspecs}. We also show the ratio of the master spectrum
from each night compared with the master spectrum from 2016 September 19 in the middle row (see \autoref{eq:strans}).
The middle rightmost panel shows the 2016 September 19 spectrum
compared with the 2013 and 2015 master spectra. The equivalent width measurement $W_{H\alpha}$ (see
\autoref{eq:wlambda}) for the $F_i/F_{Sep19}-1$ spectra are given in the bottom left of the middle panels, where
negative $W_{H\alpha}$ indicates a deeper H$\alpha$ core relative to 2016 September 19, i.e., 
less core emission. The night-to-night $W_{H\alpha}$ values can be compared to the intra-night 
changes shown in \autoref{fig:allphase} and \autoref{fig:iphase}. We also show histograms of
the transmission spectra for $50$ km s$^{-1}$ $< |v| < 200$ km s$^{-1}$ in the bottom
panels. The same comparison for the \ion{Fe}{1} control line is shown in \autoref{fig:feispecs}.

\section{H$\alpha$ TIME SERIES ANALYSIS}
\label{sec:timeseries}

\autoref{fig:haspecs} shows that all of the July nights exhibit less H$\alpha$ core emission than the 2016 September 19 observations, suggesting that HD
189733 was in a more active state, or the visible hemisphere was more active, during the night of
2016 September 19. There is some fluctuation between the July nights with 2016 July 30 showing the most core emission
and 2016 July 29 and 2016 August 01 showing the least. The 2013 data shows a similar activity level to 2016 July 30 while the 2015
data shows a much more active state than any of the other dates, with $\sim$10\% more H$\alpha$ 
core emission when compared with 2016 August 01. In a forthcoming paper, we note that the 2015 observations are 
over-active compared to the other examined nights from that study. The additional data presented here strengthens the conclusion that 
HD 189733 was especially active at that time.

In order to produce a time series of the relative changes in the H$\alpha$ core within an individual night, we produce a
``transmission'' spectrum that is identical to the definition of $S_T$ from \citet{cauley15,cauley16}:

\begin{equation}\label{eq:strans}
S_T=\frac{F_{i}}{F_{comp}}-1
\end{equation}

\noindent where $F_i$ is a single observation and $F_{comp}$ is the master comparison
spectrum constructed from the three highest signal-to-noise spectra from each night. We then 
calculate the equivalent width of the individual transmission spectra according to

\begin{equation}\label{eq:wlambda}
W_{H\alpha} = \sum\limits_{v=-200}^{+200} \left(1-\frac{F_v}{F_v^{comp}} \right) \Delta\lambda_v \ \text{\AA}
\end{equation}

\noindent where $F_v$ is the flux in the spectrum of interest at velocity $v$, $F_v^{comp}$ is the flux in the
comparison spectrum at velocity $v$, and $\Delta\lambda_v$ is the wavelength difference at velocity
$v$. Uncertainties in $W_{H\alpha}$ are calculated by summing the transmission spectrum flux
errors in quadrature across the same velocity range.

\begin{figure*}[t]
   \centering
   \includegraphics[scale=.6,clip,trim=0mm 40mm 15mm 40mm,angle=0]{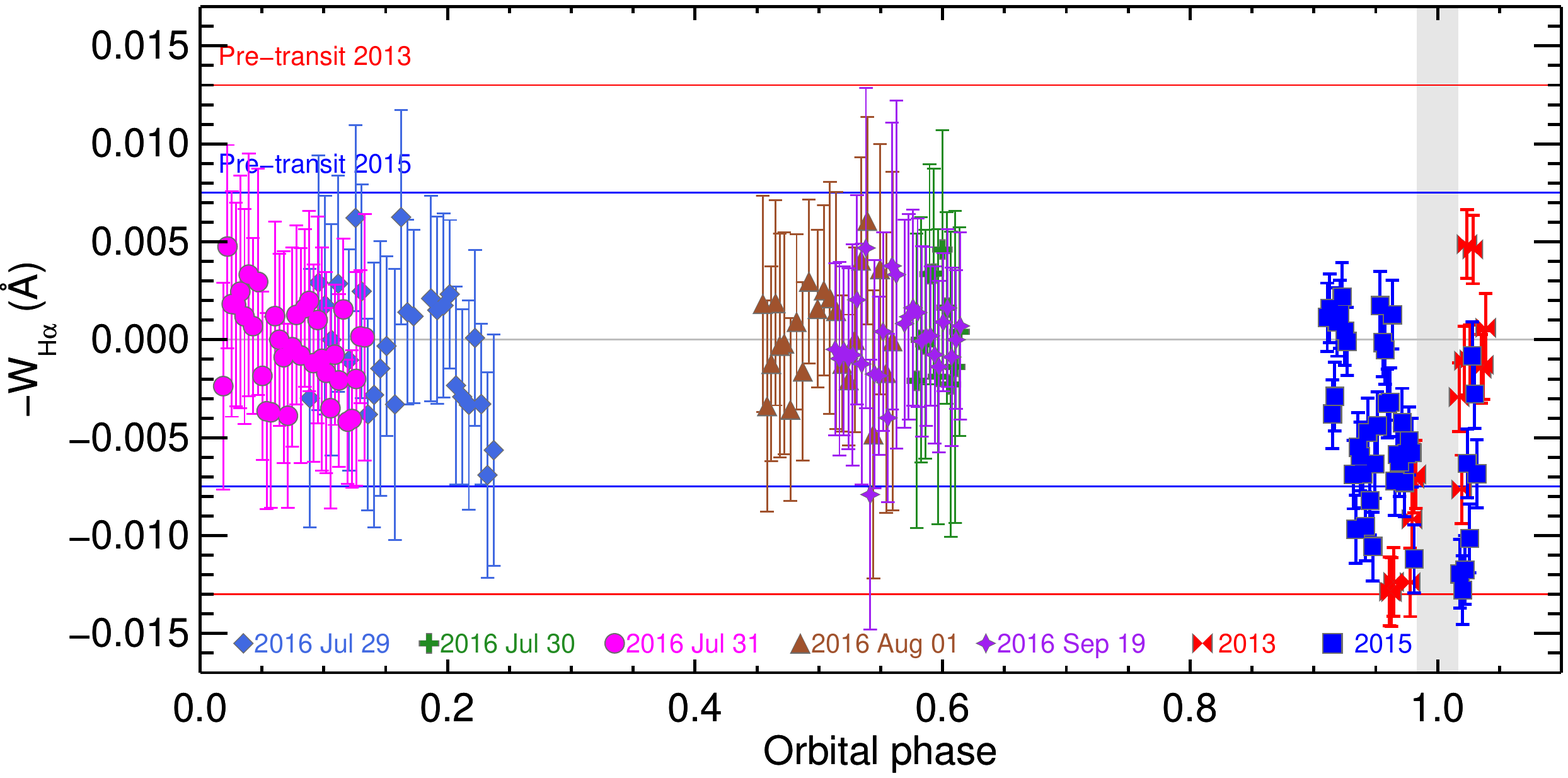} 
   \figcaption{Phase-folded $W_{H\alpha}$ data from all nights. The horizontal red and blue lines indicate
   the magnitude of the pre-transit signals measured in the 2013 and 2015 data, respectively, from \citet{cauley15,cauley16}.
   The 2013 and 2015 data sets are shown in red bowties and blue squares, respectively, near phase $\sim$0.95. 
   The transit is marked by the shaded gray region.
   Error bars represent 1$\sigma$ uncertainties derived by adding the normalized flux errors in quadrature. No changes
   in $W_{H\alpha}$ are seen similar in magnitude to the 2013 pre-transit signal. Some sporadic changes similar to the level
   of 2015 pre-transit absorption are seen but their duration is typically much shorter. \label{fig:allphase}}
\end{figure*}

\begin{figure*}[t]
   \centering
   \includegraphics[scale=.6,clip,trim=0mm 40mm 15mm 40mm,angle=0]{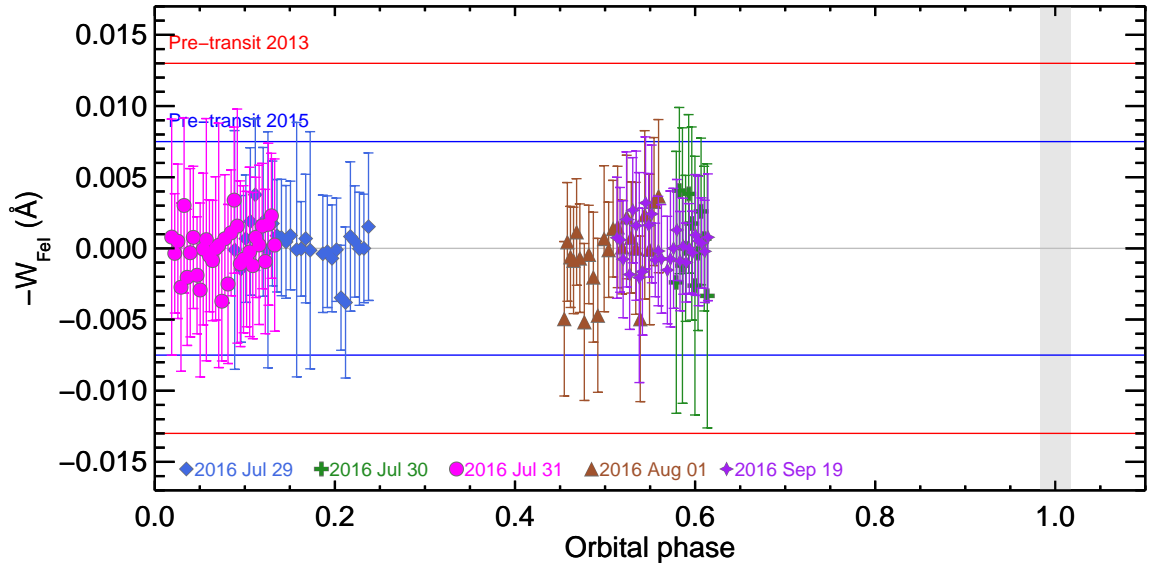} 
   \figcaption{Same as \autoref{fig:allphase} but for the \ion{Fe}{1} control line. There is less scatter in the \ion{Fe}{1} 
   measurements due to the higher signal to noise in the order and the weak telluric spectrum. Measurements of
   the \ion{Fe}{1} line are not available from the 2013 and 2015 data sets. \label{fig:allphase_cai}}
\end{figure*}

\begin{figure}[t]
   \centering
   \includegraphics[scale=.4,clip,trim=20mm 10mm 10mm 0mm,angle=0]{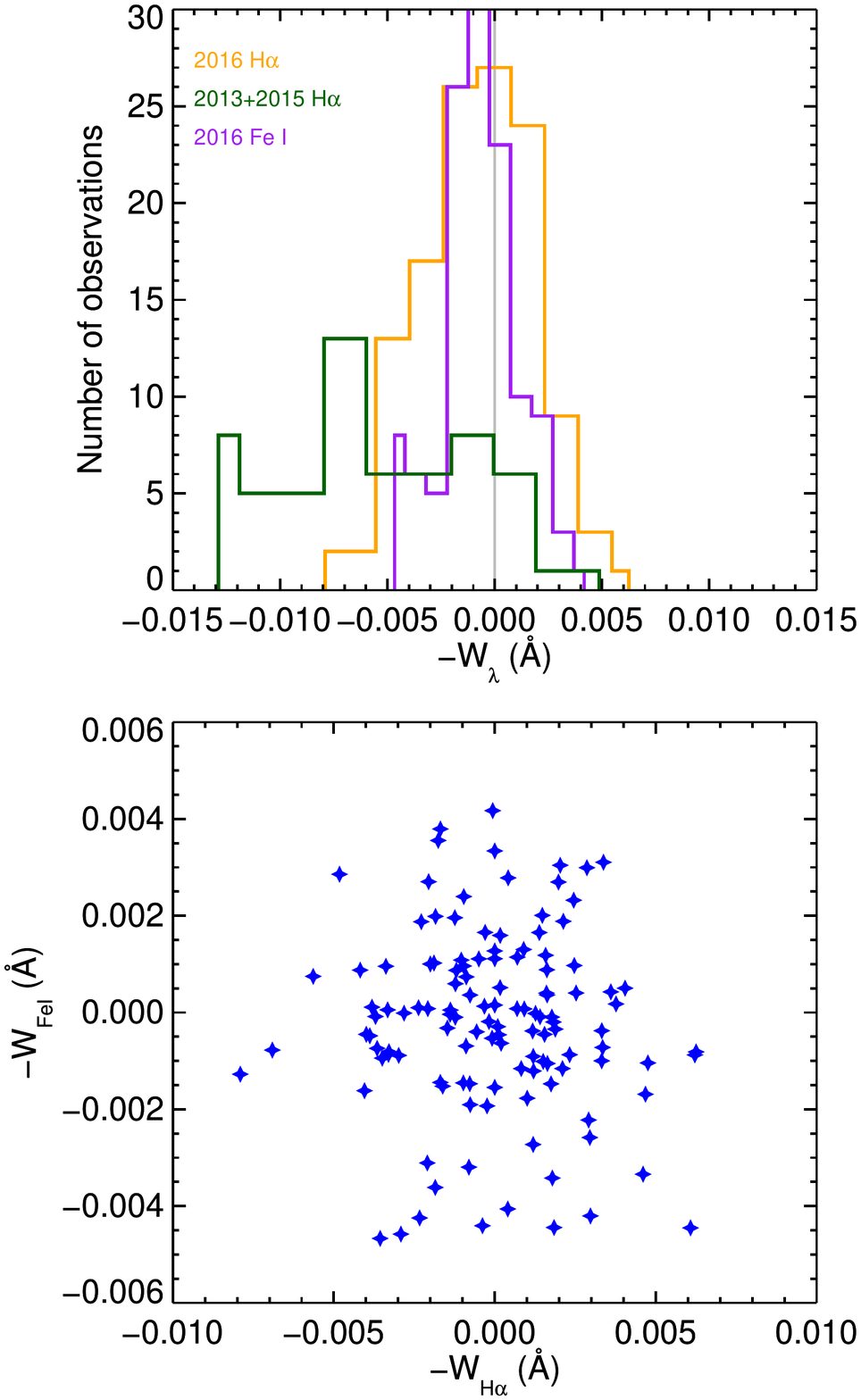} 
   \figcaption{Histogram comparison (top panel) of the 2016 out-of-transit $W_{H\alpha}$ measurements (orange line) and 
   the combined 2013 and 2015 near-transit measurements (dark green line). The \ion{Fe}{1} control line
   measurements are also shown (purple line). The 2016 data for both H$\alpha$ and \ion{Fe}{1} are approximately
   normal, with the H$\alpha$ distribution having mean $\sim$0 and FWHM$\sim$0.008, suggesting there is 
   no significant signal in the timeseries. The 2013+2015 distribution shows no identifiable structure. The bottom
   panel shows $W_{H\alpha}$ plotted against $W_{FeI}$. There is no correlation between the measurements,
   suggesting that systematics related to the extraction and transmission spectrum procedures are not
   contributing to the measured changes in $W_{H\alpha}$. \label{fig:allphase_hist}}
\end{figure}

\begin{figure*}[t]
   \centering
   \includegraphics[scale=.8,clip,trim=5mm 10mm 5mm 0mm,angle=0]{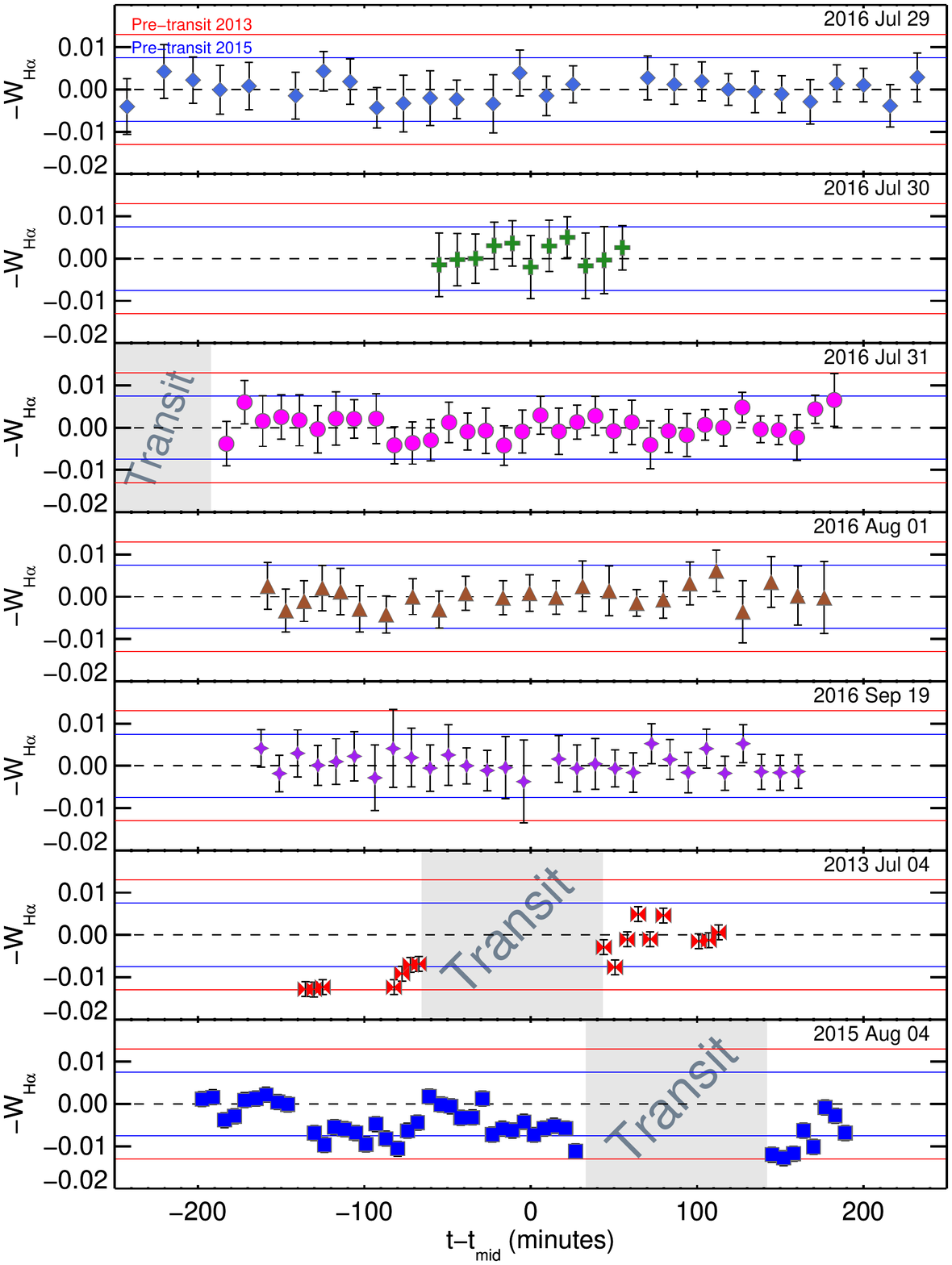} 
   \figcaption{$W_{H\alpha}$ plotted as a function of time from the midpoint (\textit{t-t$_{mid}$}) of each set of observations. Colors and
   symbols are the same as \autoref{fig:allphase}. All nights are plotted on the same vertical and horizontal
   scale. Again, no changes in $W_{H\alpha}$ similar to the 2013 pre-transit change are seen in any of the
   time series. Some variations similar to the 2015 pre-transit change are seen but their duration is shorter
   and more erratic than the two dips seen in the 2015 data. \label{fig:iphase}}
\end{figure*}

The phase-folded H$\alpha$ time series for all nights is shown in \autoref{fig:allphase} and
the \ion{Fe}{1} time series is shown in \autoref{fig:allphase_cai}. A histogram comparison
of $W_{H\alpha}$ between the 2016 dates and the combination of the 2013 and 2015 dates is shown in
 \autoref{fig:allphase_hist}. The out-of-transit points from the 2013 and 2015 data sets are also shown in \autoref{fig:allphase}. 
 The magnitude of the pre-transit signals measured in \citet{cauley15,cauley16} are marked with red and blue solid
horizontal lines. It is clear that no changes in the out-of-transit monitoring data are of
comparable strength to the 2013 pre-transit signal. Some changes, however, are of similar magnitude
to the pre-transit absorption signals from 2015 presented in \citet{cauley16}. The left panel of \autoref{fig:allphase_hist} demonstrates
that the 2016 $W_{H\alpha}$ measurements are consistent with the random measurement errors
and show no significant deviations from zero. The right panel of \autoref{fig:allphase_hist} shows that
there is no correlation between $W_{H\alpha}$ and $W_{FeI}$, suggesting that the extraction
and transmission spectrum procedures are not significantly contributing to the measurements. The same $W_{H\alpha}$ data is
shown in \autoref{fig:iphase} but as a function of time from the midpoint of each set of observations.

\begin{figure*}[t]
   \centering
   \includegraphics[scale=.65,clip,trim=5mm 25mm 10mm 60mm,angle=0]{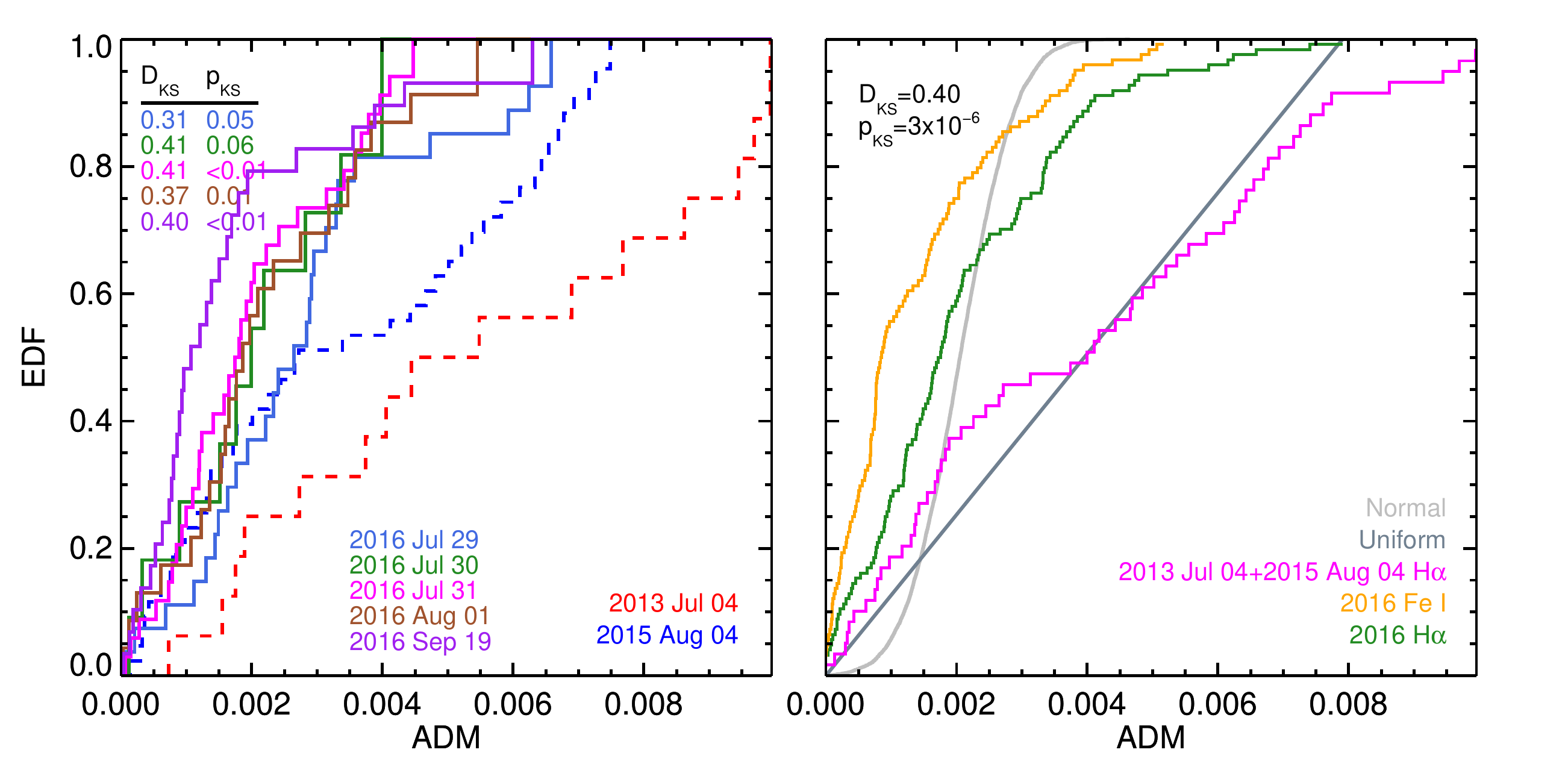} 
   \figcaption{Left panel: Empirical distribution functions (EDF) for the ADM of each time series
   compared with the EDFs from the 2013 (red dashed line) and 2015 (blue dashed line) out-of-transit
   data. With the exception of July 30, all of the EDFs differ from the 2015 EDF at the $\gtrsim 95\%$ level according to the two sided
   KS test. The KS statistics are shown in the upper left of the figure. Right panel: Total EDFs for all
   2016 H$\alpha$ observations (green line) and the combined 2013 and 2015 H$\alpha$ data (magenta line). 
   The \ion{Fe}{1} EDF is also shown (orange line), which is very similar in structure compared
   with the 2016 H$\alpha$ EDF but scaled to lower values of the ADM. Uniform
   and normal distributions for the 2016 H$\alpha$ observations are shown for reference. The combined
   EDFs differ at the $>99\%$ level, providing strong statistical evidence that the near-transit
   observation differ significantly from the far-from-transit observations in 2016. \label{fig:edfs}}
\end{figure*}

To quantify the differences in the measured $W_{H\alpha}$ time series, we have calculated the
absolute deviation from the median, or the ADM, for each night:

\begin{equation}\label{eq:adm}
ADM = |W_{H\alpha}^i-Median(W_{H\alpha})| \ \text{\AA}
\end{equation}

\noindent where $W_{H\alpha}^i$ are the individual observations and Median($W_{H\alpha}$) is the median
of all observations. The empirical distribution function (EDF) of the ADM for each night is shown in the left
panel of \autoref{fig:edfs}. The right panel of \autoref{fig:edfs} shows the total EDFs for the 2016
data (green line) and the 2013 data combined with the 2015 data (magenta line). We also include 
normal and uniform distributions, generated using the 2016 data, for reference. 
We do not make any statistical comparisons using only the 2013 data due to the smaller number
of observations compared with the 2015 data set. The 2013 and 2015
EDFs are plotted with dashed red and blue lines, respectively, in the left panel. Also shown in the left panel of 
\autoref{fig:edfs} are the two sided Kolmogorov-Smirnov (KS) statistic $D_{KS}$ and the probability
$p_{KS}$ that the null hypothesis is true, i.e., that each EDF is drawn from the same parent distribution,
for the individual 2016 EDFs compared with the 2015 EDF. The KS statistics are shown in
the right panel for the comparison between both total EDFs. 

For our purposes the EDF of the absolute deviation from the median highlights differences between a time series with structure, such as
the 2015 data, and the apparently random deviations of a time series like those from the 2016 nights. 
\autoref{fig:edfs} shows significant differences between the 2015 EDF and the 2016
EDFs: the 2015 EDF flattens off near $ADM \sim 0.0025$ and then begins to increase again near
$ADM \sim 0.0045$. In comparison, all of the 2016 EDFs are $\sim$80\% comprised of
$ADM \lesssim 0.004$. These differences are born out in the KS statistics: with the exception of the
2016 Jul 29 observations, which contain only eleven data points, all of the EDFs differ from the 2015
EDF at the 95\% level, i.e., $P_{KS}<0.05$. We can rule out the null hypothesis even more strongly
($>99\%$) for the combined EDF test. This can be interpreted as the 2016 dates all showing random deviations
from the median activity level while the 2015 data has distinct, sustained features that force data
points contained in these features to be far from the median. This contributes to the EDF at the
higher ADM values. 

The significant differences between the out-of-transit 2016 monitoring and the very near-transit data
suggest that something interesting is happening immediately before and after HD 189733 b transits. There
are two possibilities: 1. the near-transit features are due to absorbing circumplanetary material; or
2. the stellar activity level, as measured in H$\alpha$, is experiencing abnormal changes 
\textit{preferentially} at near-transit times. Using the current data it is not clear which of these two
cases is true. However, we can say with some confidence that \textit{abnormal changes in the stellar
activity level occur more frequently very close to planetary transits}. In addition, the nominal
stellar activity level is much higher for the 2015 data compared with the 2013 data set, and the
2013 data shows a similar activity level to the 2016 observations, yet both the 2013 and 2015 
observations show significantly different ADM distributions compared with the 2016 data. Thus
we cannot attribute the abnormal near-transit $W_{H\alpha}$ measurements to high levels
of stellar activity. Taken together, we believe this is strong evidence in support of SPIs
or absorption by circumplanetary material in the HD 189733 system 
\citep{cuntz,shkolnik08,shkolnik13,strugarek,miller15,pillitteri15}. 

Finally, we note that if the H$\alpha$ variations are the result of SPIs near transit, the affected region on the
star must be very near the sub-stellar point. We speculate that this is evidence for the
circumplanetary absorption interpretation since there is nothing special about the transit
in the SPI interpretation. We leave this suggestion to further investigation.

\section{SUMMARY AND CONCLUSIONS}
\label{sec:summary}

We have presented new out-of-transit, high-cadence H$\alpha$ monitoring of HD 189733 over the course
of 5 nights using the Tull Coud{\' e} spectrograph on the 2.7 meter Harlan J. Smith telescope at McDonald
Observatory with the goal of establishing the frequency of changes in the stellar activity level
at a similar magnitude to what we presented in \citet{cauley15,cauley16}. We do not find any variations
in the H$\alpha$ core flux similar to what was observed immediately pre-transit in \citet{cauley15}. 
With the exception of the 2016 July 30 data set, which only contains 11 observations, we find statistically
significant differences between each of the out-of-transit nights and the out-of-transit signal from
\citet{cauley16}. This conclusion is strengthened when the combined out-of-transit data set is
compared with the combined near-transit observations. Our results suggest that changes in 
H$\alpha$ similar to those from \citet{cauley15} and \citet{cauley16}
occur infrequently when the planet is far from transit. This is evidence for attributing the pre-transit
signals to either absorbing circumplanetary material or some type of magnetic or tidal SPI near
the sub-planetary point on the stellar surface. Further monitoring is necessary to understand the frequency
and physical nature of the near-transit changes.
 
{\bf Acknowledgments:} The authors thank the referee for their suggestions, which helped to significantly improve
this manuscript. This paper includes data taken at The McDonald Observatory of The University of 
Texas at Austin. P. W. C. is grateful to Dave Doss and Brian Roman for their instrumental knowledge
and observational support. This work was completed with support from the National Science
Foundation through Astronomy and Astrophysics Research Grant AST-1313268 (PI: S.R.).
A. G. J. is  supported by NASA Exoplanet Research Program grant 14-XRP14 2- 0090 to the University of Nebraska-Kearney.
This work has made use of NASA's Astrophysics Data System. 


\end{document}